\documentstyle[aps,preprint]{revtex}

\title{Higher-Derivative Massive Fermion Theories}
\author{ \sc{C. G. Carvalhaes$^{1,2}$, L. V. Belvedere$^1$, R. L. P. G. Amaral$^1$ {\small and} N. A. Lemos$^1$} }
\address{
$1$\small{ \it{Instituto de F\'\i sica, Universidade Federal
Fluminense\\ Av. Litor\^anea, s/n, Boa Viagem, Niter\'oi, CEP:             24210-340, RJ, Brasil.} }\\
$2$\small{ \it{Instituto de Matem\'atica e Estat\'\i stica -
Universidade do Estado do Rio de Janeiro\\ Rua S\~ao Francisco Xavier,     524, Maracan\~a, CEP: 20559-900, Rio de Janeiro, Brasil} }
}

\begin{document}

\draft

\maketitle

\begin{abstract}
We propose and canonically quantize a generalization of the two-dimensional massive fermion theory described by a Lagrangian containing third-order derivatives. In our approach the mass term contains a derivative coupling. The quantum solution is expressed in terms of three usual fermions. Employing the standard bosonization scheme, the equivalent boson theory is derived. The results obtained are used to solve a theory including a current-current interaction.
\end{abstract}

\newpage

\section{Introduction}

There is a continuing interest in quantum field theories defined by higher-derivative Lagrangians \cite{Batlle88}. In spite of their possible shortcomings, such as ghost states and unitarity violation, field theories whose equations of motion are of order higher than the second are useful to regularize ultraviolet divergences \cite{Boulware84a}, especially for supersymmetric gauge theories \cite{Fayet77}.

The appearance of curvature-squared terms as corrections to the Einstein-Hilbert Lagrangian in the effective action of superstring theories \cite{Boulware84} is a further reason why higher-derivative field theories are worth investigating for their own sake, and as such they have been studied from several different points of view in the last few years. Recently, a higher-derivative generalization of the two-dimensional free fermion theory \cite{Amaral93,Belvedere95} has been constructed and exactly solved 
by expressing the fermion fields of the model in terms of boson fields (``bosonization''). It turns out that the fermion fields that solve the higher-order equations of motion can be written in terms of usual Dirac fields, the so-called ``infrafermions''. Some of these infrafermions, however, need to be quantized with a negative metric, giving rise to an indefinite-metric Hilbert space.

In this paper we study the effect of the inclusion of a mass term on the behavior of these generalized fermion theories. We find that the requirements of Lorentz invariance, absence of tachyon excitations and hermiticity fix the form of the mass term, which differs from the usual one by the appearance of derivative couplings. The model is solved exactly and it so happens that the higher-derivative fermion fields admit only a nonlocal mapping from usual fermion fields. With the help of the standard bosoniza
tion technique \cite{Mandelstam75,Halpern75}, the solution is expressed in terms of a sine-Gordon field and of two massless free scalar fields. These results are then employed to solve a theory with a current-current interaction.

The paper is organized as follows. In Sec II we propose and canonically quantize the third-order massive fermion theory. Section III is devoted to find the equivalent-boson theory. A theory including a current-current interaction is discussed in Sec. IV. Section V is dedicated to general comments and conclusions.

\section{Canonical Quantization}

Let us consider the Lagrangian density 
\begin{equation}
{\cal L}_0(x) = -i \overline{\xi}(x)  (\partial\!\!\!/ \partial\!\!\!/^{\dagger})^N   \partial\!\!\!/ \xi(x) + m \overline{\xi}(x) (\partial\!\!\!/ \partial\!\!\!/)^N \xi(x),
\label{lag1}
\end{equation}
where $m$ is a parameter with dimension of mass. For $N>0$ the mass term is a coupling with even-order derivatives. As it will be seen, this is the simplest mass term to be introduced in order to generalize the massive fermion theory that avoids the appearance of tachyon excitations, preserves Lorentz invariance of ${\cal L}$ and provides a Hermitian Hamiltonian.

The complexity involved here is greater than in the massless case \cite{Amaral93}, when the two spinor components decouple and are treated independently. Therefore, in order to avoid unnecessarily complicated expressions, instead of considering the fairly general form (\ref{lag1}), we shall restrict ourselves to the third-order case ($N=1$). In this case it is easy to conclude by decoupling the equations of motion that if the mass term were $m\overline{\xi}\xi$ the mass acquired would be complex. No first-
derivative term like $m\overline{\xi}\partial\!\!\!/\xi$ would respect Lorentz invariance (excluding non-local terms, like $m\overline{\xi}\sqrt{\partial\!\!\!/\partial\!\!\!/}\xi)$. The second-derivative term $m\overline{\xi}\partial\!\!\!/\partial\!\!\!/\xi$ is the only Hermitian Lorentz invariant local term suitable to be introduced.

Using the light-cone variables (the conventions used here are the same as in Ref.\cite{Amaral93}, except for the definition $x^{\pm} = \frac{x^0 \pm x^1}{2}$), the Lagrangian density can be written as
\begin{equation}
{\cal L}_0(x) = -i \xi_{(1)}^*(x)\partial_{-}^3\xi_{(1)}(x) -i \xi_{(2)}^*(x)\partial_{+}^3\xi_{(2)}(x) + m \xi_{(1)}^*(x) \Box \xi_{(2)}(x) + m \xi_{(2)}^*(x) \Box \xi_{(1)}(x).
\label{lag2}
\end{equation}
Under a Lorentz transformation $x^+ \rightarrow \lambda x^+$ and $x^- \rightarrow \lambda^- x^-$ we have $\xi_{(1,2)} \rightarrow \lambda^{\mp 3/2} \xi_{(1,2)}$. The equations of motion are 
\begin{eqnarray}
\nonumber &i& \partial_{-}^3\xi_{(1)}(x) - m \Box \xi_{(2)}(x) = 0,\\
&i& \partial_{+}^3\xi_{(2)}(x) - m \Box \xi_{(1)}(x) = 0.
\end{eqnarray}
In order to quantize the theory, we must obtain the basic Poisson brackets. In accordance with the third-order character of the Lagrangian, for the basic variables we take $\xi_{(1)},\,\, \partial_{-}\xi_{(1)}, \,\, \partial_{-}^2\xi_{(1)} + im\partial_{+}\xi_{(2)}, \,\, \xi_{(2)}, \,\, \partial_{+}\xi_{(2)}, \,\, \partial_{+}^2\xi_{(2)} + im\partial_{-}\xi_{(1)}$. The associated canonical momenta, obtained by variation of the action around the equations of motion, are $-i\partial_{-}^2\xi_{(1)}^*-m\partial_{+}\xi_{(2)}^*, \,\, i\partial_{-}\xi_{(1)}^*, \,\, -i\xi_{(1)}^*, \,\, -i\partial_{+}^2\xi_{(2)}^*-m\partial_{-}\xi_{(1)}^*, \,\, i\partial_{+}\xi_{(2)}^*, \,\, -i\xi_{(2)}^*$, respectively. Our choice for basic variables, different from the one in Ref.\cite{Amaral93}, has the advantage of providing momenta with homogenous Lorentz properties, even with the mass term.

Using these variables, a systematic quantization, carried out using either a Dirac bracket formalism if $\xi^{\dagger}$ is treated as an independent variable, or Poisson brackets if $\xi^{\dagger}$ is taken as a function of $\xi$, furnishes the following nonvanishing equal-time anticommutators:
\begin{eqnarray}
\nonumber &\{&\xi_{(1)}(x),\partial_{-}^2\xi_{(1)}^*(y)\}=\{\xi_{(2)}(x),\partial_{+}^2\xi_{(2)}^*(y)\} = -\delta(x^1-y^1),\\
\nonumber &\{&\partial_{-}\xi_{(1)}(x),\partial_{-}\xi_{(1)}^*(y)\}=\{\partial_{+}\xi_{(2)}(x),\partial_{+}\xi_{(2)}^*(y)\} = \delta(x^1-y^1),\\
\nonumber &\{&\partial_{-}\xi_{(1)}(x),\partial_{+}^2\xi_{(2)}^*(y)\}=\{\partial_{+}\xi_{(2)}(x),\partial_{-}^2\xi_{(1)}^*(y)\} = im\delta(x^1-y^1),\\
&\{&\partial_{-}^2\xi_{(1)}(x),\partial_{-}^2\xi_{(1)}^*(y)\}=\{\partial_{+}^2\xi_{(2)}(x),\partial_{+}^2\xi_{(2)}^*(y)\} = m^2\delta(x^1-y^1).
\label{anticom1}
\end{eqnarray}

Introducing the Fourier decomposition
\begin{equation}
\xi_{(\alpha)}(x) = \int d^2k e^{-ikx} \tilde \xi_{(\alpha)}(k),
\end{equation}
we obtain the general solution
\begin{eqnarray}
\nonumber\tilde\xi_{(1)}(k) &=& a(k) \delta(k^2-m^2) + b_{(2)}(k_-) \delta(k_+) + c_{(1)}(k_+) \delta(k_-) - \frac{k_+^2}{m} b_{(1)}(k_+) \frac{d}{dk_-} \delta(k_-),\\
\tilde\xi_{(2)}(k) &=& \frac{k_-^3}{m^3} a(k) \delta(k^2-m^2) + b_{(1)}(k_+) \delta(k_-) + c_{(2)}(k_-) \delta(k_+) - \frac{k_-^2}{m} b_{(2)}(k_-) \frac{d}{dk_+} \delta(k_+).
\label{sol1}
\end{eqnarray}

With the help of fields $\chi^i$ with dispersion relations described by
\begin{eqnarray}
\nonumber
&\chi^1_{(1,2)}&(x) = -i\int d^2k\, a(k)\, k_{\mp}\, \delta(k^2-m^2)\, e^{-ikx},\\
\nonumber
\frac{1}{\partial_\pm}&\chi^2_{(1,2)}&(x^\pm) = \int dk_{\pm}\, c_{(1,2)} (k_{\pm})\, e^{-ik_{\pm} x^\pm},\\
\frac{1}{\partial_\pm}&\chi^3_{(1,2)}&(x^\pm) = \int dk_{\pm}\, b_{(1,2)} \,(k_{\pm})\, e^{-ik_{\pm} x^\pm}
\label{modea}
\end{eqnarray}
and consistently defining the operators
\begin{equation}
\frac{1}{\partial_\mp}\chi^1_{(1,2)}(x) = -\frac{1}{m^2} \partial_\pm \chi^1_{(1,2)}(x),
\end{equation}
we come back to the configuration space arriving at
\begin{eqnarray}
\nonumber
\xi_{(1)}(x) &=& \frac{1}{\partial_{-}}\chi^1_{(1)}(x) + \frac{1}{\partial_{+}}\chi^2_{(1)}(x^+) + \frac{1}{\partial_{-}}\chi^3_{(2)}(x^-) + i\frac{x^-}{m} \partial_{+}\chi^3_{(1)}(x^+),\\
\xi_{(2)}(x) &=& \frac{1}{\partial_{+}}\chi^1_{(2)}(x) + \frac{1}{\partial_{-}}\chi^2_{(2)}(x^-) + \frac{1}{\partial_{+}}\chi^3_{(1)}(x^+) + i\frac{x^+}{m} \partial_{-}\chi^3_{(2)}(x^-).
\end{eqnarray}
A convenient definition to $\frac{1}{\partial_{\pm}} \chi^{2,3}_{(1,2)}$ is
\begin{equation}
\frac{1}{\partial_\pm}\chi^{2,3}_{(1,2)}(x^\pm) = \frac{1}{2}\int_{-\infty}^{x^\pm} dz^1\, \chi^{2,3}_{(1,2)}(z^\pm).
\label{sol2}
\end{equation}
However, for calculating physical quantities, other definitions could be applied so long as the identity
\begin{equation}
\partial_\pm\left\{\frac{1}{\partial_\pm}\chi_{(1,2)}(x^\pm)\right\} = \chi_{(1,2)}(x^\pm)
\end{equation}
is ensured to hold.

The mode $\chi^1$ is massive, whereas $\chi^2$ and $\chi^3$ are massless. In the general case (\ref{lag1}) this decomposition would generate one massive mode and $N-1$ other massless modes. Tachyon excitations do not appear. 

Now we proceed to find the anticommutation laws of the modes $\chi^i$. Inverting the relations (\ref{sol2}) we obtain
\begin{eqnarray}
\nonumber
\chi^1_{(1)}(x) &=& \frac{i}{m} \partial_{+}^2\xi_{(2)}(x) - \frac{2i}{m} \partial_1\partial_{+}\xi_{(2)}(x) -   \frac{2}{m^2}\partial_1\partial_{-}^2\xi_{(1)}(x),\\
\nonumber
\chi^2_{(1)}(x^+) &=& 2\partial_1\xi_{(1)}(x) + (1+\frac{4}{m^2}\partial_1^2)\partial_{-}\xi_{(1)}(x) +   \frac{4}{m^2}(\partial_1 - \frac{2}{m^2}\partial_1^3 - x^-\partial_1^2)\partial_{-}^2\xi_{(1)}(x) \\
\nonumber
&+& \frac{2i}{m}(\partial_1 - \frac{4}{m^2}\partial_1^3 - 2x^-\partial_1^2)\partial_{+}\xi_{(2)}(x) - \frac{i}{m}\partial_{+}^2\xi_{(2)}(x),\\
\chi^3_{(1)}(x^-) &=& \partial_{+}\xi_{(2)}(x) - \frac{i}{m}\partial_{+}^2\xi_{(2)}(x).
\end{eqnarray}
The lower components can be calculated likewise or by simply switching the spinor indices and $x^1$ to $-x^1$. Using the above relations and the anticommutation laws (\ref{anticom1}) one can verify, after a tedious algebra, the following nonvanishing anticommutation relations:
\begin{eqnarray}
\nonumber
&\{&\chi^1_{(\alpha)}(x),{\chi^1}^{\dagger}_{(\alpha)}(y)\} = \delta(x^1-y^1),\\
\nonumber
&\{&\chi^2_{(\alpha)}(x),{\chi^2}^{\dagger}_{(\alpha)}(y)\} = -\frac{16}{m^4} \frac{\partial^4}{\partial(x^1)^4} \delta(x^1-y^1),\\
\nonumber
&\{&\chi^3_{(\alpha)},{\chi^2}^{\dagger}_{(\alpha)}(y)\} = -\frac{2i{\gamma^5}_{\alpha\alpha}}{m}\frac{\partial}{\partial x^1}  \delta(x^1-y^1),\\
&\{&\chi^3_{(\alpha)}(x),{\chi^3_{(\alpha)}}^{\dagger}(y)\} = 0.
\label{anticom2}
\end{eqnarray}

From Eq.(\ref{modea}) one can check that
\begin{equation}
(i\partial\!\!\!/ - m)\chi^1(x) = 0.
\end{equation}
It is a straightforward exercise to verify that the dimension and Lorentz properties of $\chi^1$ are the same as those of usual fermions. Therefore $\chi^1$ is a massive Dirac field quantized with positive metric. The other two modes are noncanonical. Nevertheless, the anticommutation structure (\ref{anticom2}) can be cast into a diagonal form by introducing a free massive field $\psi^1$ and two other free massless fields $\psi^2$ and $\psi^3$ quantized with opposite metrics:
\begin{equation}
\{\psi^1(x),{\psi^1}^{\dagger}(y)\} =
\{\psi^2(x),{\psi^2}^{\dagger}(y)\} = -\{\psi^3(x),{\psi^3}^{\dagger}(y)\} = \delta(x^1-y^1).
\end{equation}
In terms of these fields, for an arbitrary integer $p$ (see Appendix), we have
\begin{eqnarray}
\nonumber
\chi^1_{(\alpha)} &=& \psi^1_{(\alpha)},\\
\nonumber
\chi^2_{(1,2)} &=& \frac{1}{2M^{p+1}} \partial_{\pm}^{p+1} (\psi^2_{(1,2)} + \psi^3_{(1,2)}) + (-1)^p \frac{M^{p+1}}{2m^4} \partial_{\pm}^{3-p} (\psi^2_{(1,2)} - \psi^3_{(1,2)}),\\
\chi^3_{(1,2)} &=& i(-1)^p \frac{M^{p+1}}{m} \partial_{\pm}^{-p} (\psi^2_{(1,2)} - \psi^3_{(1,2)}),
\label{modes3}
\end{eqnarray}
where $M$ is an arbitrary parameter of the same dimension as $m$. Under Lorentz transformations \cite{Belvedere95} we require that $M\rightarrow \lambda^{\frac{1-p}{p+1}}M$. Using this mapping, the original field turns out to be
\begin{eqnarray}
\nonumber
\xi_{(1,2)} &=& \frac{1}{\partial_{\mp}}\psi^1_{(1,2)} + \frac{1}{2M^{p+1}} \partial_{\pm}^p (\psi^2_{(1,2)} + \psi^3_{(1,2)})\\
&+& (-1)^p \frac{M^{p+1}}{m} \left[(\frac{1}{2m^3}\partial_{\pm}^{2-p} - \frac{x^{\mp}}{m} \partial_{\pm}^{1-p})(\psi^2_{(1,2)} - \psi^3_{(1,2)}) + i\partial_{\mp}^{-p-1}(\psi^2_{(2,1)} - \psi^3_{(2,1)})\right].
\label{sol3}
\end{eqnarray}
Note that it is impossible to adjust $p$ to describe the original fields locally in terms of usual fermions, while the corresponding relationship is local in the massless case \cite{Amaral93,Belvedere95}.

The conserved currents associated with the global gauge symmetry are given by the products of fields and conjugate  momenta. The light-cone components (which do not satisfy independent  conservation laws) are
\begin{eqnarray}
\nonumber
j^- &=& i\xi_{(1)}(i\partial_{-}^2\xi_{(1)}^* + m\partial_{+}\xi_{(2)}^*) + (\partial_{-}\xi_{(1)})(\partial_{-}\xi_{(1)}^*) - i(-i\partial_{-}^2\xi_{(1)} + m\partial_{+}\xi_{(2)})\xi_{(1)}^*,\\
j^+ &=& i\xi_{(2)}(i\partial_{+}^2\xi_{(2)}^* + m\partial_{-}\xi_{(1)}^*) + (\partial_{+}\xi_{(2)})(\partial_{+}\xi_{(2)}^*) - i(-i\partial_{+}^2\xi_{(2)} + m\partial_{-}\xi_{(1)})\xi_{(2)}^*.
\label{cur1}
\end{eqnarray}
Using these expressions and the diagonal expansions of $\xi$ we arrive at
\begin{equation}
j^\mu(x) = \overline{\psi^1}(x) \gamma^\mu \psi^1(x) + \overline{\psi^2}(x) \gamma^\mu \psi^2(x) - \overline{\psi^3}(x) \gamma^\mu \psi^3(x),
\label{cur2}
\end{equation}
where a surface term has been dropped out. Defining
\begin{equation}
Q = \int{ dz^1 j^0(z) },
\label{charge}
\end{equation}
it is straightforward to show from (\ref{charge}), (\ref{sol3}) and (\ref{cur2}) that
\begin{equation}
\{Q,\xi(x)\} = -\xi(x).
\end{equation}

\section{Bosonization}
 
As emphasized in \cite{Belvedere95}, the Hamiltonian ${\cal H}_0$ obtained from the Legendre transformation of the Lagrangian (\ref{lag2}) evolves the $\xi$ field. The Hamiltonian ${\cal H}^\prime_0$ evolving the infrafermions is obtained from it by recognizing the time-dependent relationship between the basic variables and the infrafermions as a point transformation. The generating function may be constructed as in \cite{Belvedere95} and the Hamiltonian ${\cal H}^\prime_0$ computed. The result is the Hami
ltonian for the three independent and canonical (except for metrics) first-derivative infrafermions: 
\begin{equation}
{\cal H}^\prime_0 = -i\overline{\psi}^1\gamma^1\partial_1\psi^1 - i\overline{\psi}^2\gamma^1\partial_1\psi^2 + i\overline{\psi}^3\gamma^1\partial_1\psi^3 +
m\overline{\psi}^1\psi^1.
\label{hhh}
\end{equation}
By means of a Legendre transformation one finds
\begin{equation}
{\cal L}^\prime(x) = \overline{\psi}^1(x)(i\partial\!\!\!/-m)\psi^1(x)+\overline{\psi}^2(x)(i\partial\!\!\!/)\psi^2(x)-\overline{\psi}^3(x)(i\partial\!\!\!/)\psi^3(x).
\label{lll}
\end{equation}
It is the Hamiltonian (\ref{hhh}) and the infrafermions that we are going to bosonize. The bosonization scheme we employ is the standard one \cite{Mandelstam75,Halpern75}. Therefore we have
\begin{equation}
\psi^j_{(\alpha)}(x) = (\frac{\mu}{2\pi})^{1/2} :e^{-i\sqrt{\pi}\{\int_{-\infty}^x dz^1 \pi_j(z) + \gamma^5_{\alpha\alpha}\phi_j(x)\}}:\,\,\, ,
\label{bos1}
\end{equation}
where $\mu$ is an arbitrary finite mass scale, $\phi_2$ and $\phi_3$ are free and massless scalar fields, $\phi_1$ is a sine-Gordon field and $\pi_j = \dot\phi_j$. In the last expression we have suppressed the Klein factors that ensure the anticommutation relations.
Its worth remarking that opposite metrics are ensured by Klein factors too \cite{Amaral93}.

Thus, the equivalent boson field theory Hamiltonian is
\begin{equation}
{\cal H}^B_0 = \frac{1}{2} [\pi^2_1 + (\partial_1\phi_1)^2] - \frac{m}{\pi}\mu cos(2\sqrt{\pi}\phi_1) + \frac{1}{2} [\pi^2_2 + (\partial_1\phi_2)^2] + \frac{1}{2} [\pi^2_3 + (\partial_1\phi_3)^2].
\label{bbb}
\end{equation}
For the conserved current (\ref{cur2}) we find
\begin{equation}
\j^\mu(x) = -\frac{1}{\sqrt{\pi}}\varepsilon^{\mu\nu}\partial_\nu \{\phi_1(x) + \phi_2(x) - \phi_3(x) \}.
\end{equation}
The bosonization of the higher-derivative fermion field is obtained by using (\ref{bos1}) in (\ref{sol3}).

\section{Current-current Interaction}

Consider the theory described by
\begin{equation}
{\cal L}_1(x) = {\cal L}_0(x) + g\,j^+(x)j^-(x),
\label{lc}
\end{equation}
where ${\cal L}_0$ is the Lagrangian density (\ref{lag2}), $j^{\pm}$ are given by (\ref{cur1}), $g$ is a constant and all the fields are in the Heisenberg picture. This is a more general third-order theory that contains a current-current interaction term. A natural candidate to be the infrafermion Lagrangian density for this theory is built by adding the current-current interaction (\ref{cur2}) to the Lagrangian density (\ref{lll}) in the Heisenberg picture. This identification is correct in the interactio
n picture, since the solution (\ref{sol3}) has led us to identify the third-order Lagrangian density (\ref{lag2}) with the first-order fermion theory (\ref{lll}) and the current (\ref{cur1}) with (\ref{cur2}). However, it is not clear that this direct identification remains in the Heisenberg picture. It depends on generalizing the solution (\ref{sol3}), an issue we do not address here. In order to gain insight into this new theory we shall then consider the first-order fermion theory
\begin{equation}
{\cal L} = \overline{\Psi}^1(i\partial\!\!\!/ - m)\Psi^1 + \overline{\Psi}^2(i\partial\!\!\!/)\Psi^2 -\overline{\Psi}^3(i\partial\!\!\!/)\Psi^3- g(\overline{\Psi}^1 \gamma^{\mu} \Psi^1 + \overline{\Psi}^2 \gamma^{\mu} \Psi^2 - \overline{\Psi}^3 \gamma^{\mu} \Psi^3)^2.
\label{ifti}
\end{equation}
From now on, we shall use lower case letters to denote fields in the interaction picture and the upper case ones to those in the Heisenberg picture. We have been led to a Thirring model with global $SU(2,1)$ symmetry explicitly broken by the mass term.

Following \cite{Halpern75} in the interaction picture the current-current term (\ref{cur2}) should be written as
\begin{equation}
{\cal H}^B_I = \frac{g}{2}\left[(j_F^0)^2 - \lambda(j_F^1)^2\right],
\end{equation}
where $\lambda$ is a parameter that has to be introduced in the interaction picture and is fixed by requiring Lorentz invariance. The subscript $F$ was inserted in order to emphasize that $j^\mu$ is a functional of free quantities, since we are in the interaction picture. After bosonization we find that
\begin{equation}
{\cal H}_I = \frac{g}{2} \left\{  (\partial_1\phi_1 + \partial_1\phi_2 - \partial_1\phi_3)^2 - \lambda (\pi_1 + \pi_2 - \pi_3)^2 \right\}.
\end{equation}
The full Heisenberg picture bosonized Hamiltonian density is immediately found:
\begin{eqnarray}
\nonumber {\cal H} &=& {\cal H}^\prime_0[\Phi,\Pi] + {\cal H}_I[\Phi,\Pi] = \frac{1}{2} \left[\Pi_1^2 + (\partial_1\Phi_1)^2\right] - \frac{m}{\pi}\,\mu\, cos(2\sqrt{\pi}\Phi_1) + \frac{1}{2} \left[\Pi_2^2 + (\partial_1\Phi_2)^2\right]\\
&+& \frac{1}{2} \left[\Pi_3^2 + (\partial_1\Phi_3)^2\right] + \frac{g}{2}(\partial_1\Phi_1 + \partial_1\Phi_2 - \partial_1\Phi_3)^2  - \frac{g\lambda}{2} (\Pi_1  + \Pi_2 - \Pi_3)^2.
\end{eqnarray}
It is a simple exercise to show that the Heisenberg picture momenta $\Pi_i$ are
\begin{eqnarray}
\nonumber
\Pi_1 &=& \frac{1-2g\lambda}{1-3g\lambda}\dot\Phi_1 + \frac{g\lambda}{1-3g\lambda}\dot\Phi_2 - \frac{g\lambda}{1-3g\lambda}\dot\Phi_3,\\
\nonumber
\Pi_2 &=& \frac{g\lambda}{1-3g\lambda}\dot\Phi_1 + \frac{1-2g\lambda}{1-3g\lambda}\dot\Phi_2 - \frac{g\lambda}{1-3g\lambda}\dot\Phi_3,\\
\Pi_3 &=& - \frac{g\lambda}{1-3g\lambda}\dot\Phi_1 - \frac{g\lambda}{1-3g\lambda}\dot\Phi_2 + \frac{1-2g\lambda}{1-3g\lambda}\dot\Phi_3.
\label{PPP}
\end{eqnarray}
The fields in the two pictures are related by $A_H = U^\dagger A_I U,\,\, \dot U = -i {\cal H}_I U$.

A Legendre transformation yields the full Heisenberg picture Lagrangian $\cal L$. Requiring Lorentz invariance of $\cal L$ we obtain $\lambda = \frac{1}{1+3g}$. This result could also be achieved by imposing Schwinger's condition \cite{Halpern75}. Thus,
\begin{eqnarray}
\nonumber {\cal L} &=& \frac{1+g}{2}(\partial_\mu\Phi_1)^2 + \frac{1+g}{2}(\partial_\mu\Phi_2 )^2 + \frac{1+g}{2}(\partial_\mu\Phi_3)^2 +  g(\partial_\mu\Phi_1)(\partial^\mu\Phi_2)\\
&-& g(\partial_\mu\Phi_1)(\partial^\mu\Phi_3)-
g(\partial_\mu\Phi_2)(\partial^\mu\Phi_3) + \frac{m}{\pi}\,\mu\, cos(2\sqrt{\pi}\Phi_1).
\label{lag3}
\end{eqnarray}

The transformations
\begin{eqnarray}
\nonumber
\Phi_1 &=& \frac{\sqrt{1+2g}}{\sqrt{1+3g}} \Phi_1^\prime,\\
\nonumber
\Phi_2 &=& \frac{-g}{\sqrt{(1+2g)(1+3g)}} \Phi_1^\prime + \frac{\sqrt{1+g}}{\sqrt{1+2g}} \Phi_2^\prime,\\
\Phi_3 &=& \frac{g}{\sqrt{(1+2g)(1+3g)}} \Phi_1^\prime + \frac{g}{\sqrt{(1+g)(1+2g)}} \Phi_2^\prime + \frac{1}{\sqrt{1+g}} \Phi_3^\prime
\label{diag2}
\end{eqnarray}
applied to $\cal L$ leave it diagonal, i.e.,
\begin{equation}
{\cal L} = \frac{1}{2}(\partial_\mu\Phi_1^\prime)^2 + \frac{1}{2}(\partial_\mu\Phi_2^\prime)^2 + \frac{1}{2}(\partial_\mu\Phi_3^\prime)^2 + \frac{m}{\pi}\mu cos(2\sqrt{\pi}a\Phi_1^\prime),
\label{lag4}
\end{equation}
where $a$ is the coefficient of $\Phi_1^\prime$ in the first of Eqs.(\ref{diag2}). 

Having obtained the canonical scalar fields, let us derive the bosonized expression of the infrafermions in the Heisenberg picture. It amounts to writing all operators in (\ref{sol3}) as Heisenberg field operators. This means applying the transformations (\ref{PPP}) and (\ref{diag2}) on
\begin{equation}
\Psi^j_{(\alpha)}(x) = (\frac{\mu}{2\pi})^{1/2} :e^{-i\sqrt{\pi}\{\int_{-\infty}^x dz^1 \Pi_j(z) + \gamma^5_{\alpha\alpha}\Phi_j(x)\}}:.
\label{bos2}
\end{equation}
The dynamics of the fields $\Phi^{\prime}_i$ is found from (\ref{lag4}). From (\ref{bos2}) and (\ref{lag4}) all expected values of infrafermion fields may be computed. It is worthwhile to comment that in the general case (\ref{lag1}) one would be led to a Thirring model with $SU(N+1,N)$ explicitly broken global symmetry

\section{Conclusion}

We have discussed here the generalization of the massive fermion theory by introducing higher derivatives. The requirements of Lorentz symmetry, hermiticity of the Hamiltonian, and absence of tachyon excitations suffice to fix the mass term. The mode expansion of the fermion fields has been explicitly made and it has been seen that one needs two massless first-order (infra) fermion fields and one massive field to express the solution in familiar terms. In contrast to the massless case the relation between 
the higher-derivative field and the infrafermions is non-local. A family of (equivalent) solutions has been constructed but all of them are non-local in some degree. The interesting point is that, in spite of the non-local relationship among the fields, the current expressed in terms of the infrafermions becomes the sum of the currents associated with each infrafermion including the negative sign expected for the negative metric infrafermion.

As an example of application we have bosonized the model resulting from the current-current interaction expressed in terms of the infrafermions. The new infrafermion fields have been obtained, what allows the computation of any number of correlation functions. The bosonized model is written in terms of one massive and two massless scalars. The effect of the interaction appears in the change in the value of the mass and in the dependence of the infrafermions in all scalar fields.

The generalization of the model by considering coupling with a gauge field, as in Ref.\cite{Amaral93}, is presently under investigations. Due to the presence of derivatives on the mass term this generalization is not a trivial rewriting of the treatment of the massive Schwinger model.

{\small \noindent{\underline{\bf \it Acknowledgments}} : The authors express their thanks to Conselho Nacional de Desenvolvimento Cient\'{\i}fico e Tecnol\'ogico (CNPq) and to Coordena\c c\~ao de Aperfei\c coamento de Pessoal de N\'\i vel Superior (CAPES), Brazil, for partial financial support.}

\appendix
\section*{\bf Diagonalization}

For the sake of simplicity, we shall concentrate only on the first component and derive the case $p=1$. The extensions can be easily obtained by just following the same procedure.

Our first step consists in finding combinations $\alpha$ and $\beta$ of $\chi^2_{(1)}$ and $\chi^3_{(1)}$ such that
\begin{eqnarray}
\nonumber &\{&\alpha(x^+) , \alpha^*(y^+)\} = 0,\\
\nonumber &\{&\beta(x^+) , \beta^*(y^+)\} = 0,\\
&\{&\alpha(x^+) , \beta^*(y^+)\} = \delta(x^1-y^1).
\label{anticomap}
\end{eqnarray}

From (\ref{anticom2}) we obtain
\label{anticomap2}
\begin{eqnarray}
\nonumber
&\{&\frac{1}{\partial_{+}}\chi^2_{(1)}(x^{+}),\frac{1}{\partial_{+}}{\chi^2_{(1)}}^{\!\!\! *}(y^{+})\} = \frac{4}{m^4}\partial_1^2\delta(x^1-y^1),\\
&\{&\partial_{+}\chi^3_{(1)}(x^{+}),\frac{1}{\partial_{+}}{\chi^2_{(1)}}^{\!\!\! *}(y^{+})\} = -\frac{i}{m}\partial_1^2\delta(x^1-y^1).
\end{eqnarray}
Defining 
\begin{eqnarray}
\nonumber
\alpha(x^{+}) &=& M\frac{1}{\partial_{+}}\chi^2_{(1)}(x^{+}) + b\,\,\partial_{+}\chi^3_{(1)}(x^{+}),\\
\beta(x^{+}) &=& c\,\,\partial_{+}\chi^3_{(1)}(x^{+}),
\end{eqnarray}
the relations (\ref{anticomap}) are supplied by taking 
\begin{equation}
b = i\frac{M}{2m^3} ,\,\,\,c = i\frac{m}{M}.
\end{equation}
Now we have only to adjust $M$ to get $\psi^2_{(1)}$ and $\psi^3_{(1)}$ from the combinations $\alpha + \beta$ and $\alpha - \beta$, respectively. Inverting these relations we obtain (\ref{modes3}) for $p=1$.

\end{document}